# Empirical Evidence for the Relevance of Fractional Scoring in the Calculation of Percentile Rank Scores

**Michael Schreiber**

*Institute of Physics, Chemnitz University of Technology, 09107 Chemnitz, Germany. E-mail: schreiber@physik.tu-chemnitz.de*

Fractional scoring has been proposed to avoid inconsistencies in the attribution of publications to percentile rank classes. Uncertainties and ambiguities in the evaluation of percentile ranks can be demonstrated most easily with small datasets. But for larger datasets an often large number of papers with the same citation count leads to the same uncertainties and ambiguities which can be avoided by fractional scoring. This is demonstrated for four different empirical datasets with several thousand publications each which are assigned to 6 percentile rank classes. Only by utilizing fractional scoring the total score of all papers exactly reproduces the theoretical value in each case.

## Introduction

Leydesdorff, Bornmann, Mutz, and Opthof (2011) proposed percentile-based indicators for the evaluation of publications based on their position within a given citation distribution. The basic idea is to assign publications to a percentile rank (PR) class and then attribute a weight according to the PR class to determine the score of the publication. There is no unique way of appointing papers to PR classes and different suggestions have been presented (Hyndman & Fan, 1996; Sheskin, 2007; Leydesdorff et al., 2011; Rousseau, 2012; Pudovkin & Garfield, 2009). All these proposals can lead to inconsistencies in the behavior of the calculated citation impact indicators (Schreiber, 2012b). Leydesdorff and Bornmann (2012) are afraid that the discussion "may have opened a box of Pandora allowing for generating a parameter space of other possibilities".

Most of the problems are created by assigning the same PR class and thus the same (usually integer) weight to a large number of papers with the same citation count. A small change in the citation distribution can then shift all these tied papers from one PR class to the other and thus produce large changes in the scoring (Schreiber, 2012b; Waltman & Schreiber, 2013). I have recently proposed to average the weights of the tied papers and assign the average weight to all tied papers (Schreiber, 2012b). For several example sets with a small number of publications it was demonstrated that in this way the mentioned inconsistencies can be avoided nearly completely. A small remaining imperfection could be traced to the discretization of the PR classes. Therefore I suggested the fractional scoring (Schreiber, 2012b) as a new scoring rule as the final solution of the problem. In this fractional scoring scheme the publications at the border between two different PR classes are shared between the respective PR classes and attributed fractional weights corresponding to their shares. This approach has been elaborated and applied for the calculation of the indicator counting the top 10% most frequently cited papers (Waltman et al., 2013). There it was shown in a formal mathematical framework that this fractional scoring exactly reproduces the theoretical value for the total score of all papers. Some simple examples again for a small number of papers have been analyzed (Schreiber, 2012c) showing that the previously criticized uncertainties and ambiguities in the evaluation of PR classes do not occur in the fractional scoring scheme.

In a recent letter Bornmann (2012) has discounted the fractional scoring approach and other procedures for the calculation of PR scores, claiming that "the differences between the various methods of the PR score calculation proposed might be of little and no practical consequence" when using large datasets. Indeed the previously presented examples (Schreiber, 2012b, 2012c; Waltman et al., 2013) comprised small numbers of publications; this, however, was due to the fact that the problems and their solution could be demonstrated most clearly by utilizing such small datasets. Waltman et al. (2013) mentioned already that these problems and their solution are of empirical relevance also for large datasets of several ten thousand publications, because a sizable number of publications of tied papers occurred at the threshold of the top 10% most frequently cited publications in various fields. It is the purpose of the present investigation to demonstrate these problems quantitatively for different empirical datasets thus showing the relevance of fractional scoring.

## Different scoring rules and their application to the first example

*PR classes and thresholds*

All evaluations below are performed for the case of 6 PR classes yielding the $I3(6)$ indicator for the total score or, respectively, the $R(6) = I3(6)/N$ indicator for the relative score, normalizing $I3$ with the total number $N$ of publications in the dataset, as proposed by Bornmann and Mutz (2011). The 6 PR classes distinguish the bottom 50%, 50%-75%, 75%-90%, 90%-95%, 95%-99%, and top 1% publications determined according to the number of citations which the papers received in a given time interval.

As a first example I have evaluated the publication data of 26 physicists from my home Institute of Physics at Chemnitz University of Technology, which I had previously investigated (Schreiber, 2008a, 2009, 2010a). The data have been collected from the Web of Science in January and February 2007 and comprise 2373 publications with a total number of 25554 citations. This is considered here as the reference set for an evaluation of the individual scientists. For the purpose of the present study it is sufficient to analyze the complete set without caring about the attribution of the papers to the different researchers. But the idea is that these researchers can be evaluated in comparison with this reference set. Usually a larger reference set is used. Nevertheless the present approach refers only to papers of the same field (physics) and the same document type (article) and thus provides a reasonably homogeneous sample. In principle one should further distinguish different publication years in the reference set. This is not realized in the present investigation, because then the reference sets per year would become rather small. However, as the purpose of the present study is not evaluation and comparison of the individual scientists, but rather to show the significance of tied publications in the application of different scoring schemes, the distinction of the publication years is not urgent,

The boundaries $p_k$ of the above mentioned 6 percentage intervals are listed in Table 1; after sorting the publications according to their numbers of citations one can easily determine the numbers of citations which the publications at these thresholds have received. These numbers are also given in Table 1, as well as the numbers of publications below, at, and above the thresholds. It is well known that the citation distributions are usually strongly skewed. Therefore it is not surprising that the 50% threshold is already reached with only 4 citations. 126 publications with 4 citations occur in the dataset, which means that we have 5.31% of all publications at this threshold, see also Table 1. At the 75% boundary we find 50 publications with 12 citations each, i.e., 2.10% of all publications are exactly at the threshold. But even at the 90% boundary where 25 citations are needed, we still have 9 tied papers at the threshold, i.e., 0.37%. These numbers indicate already the problem, namely how to assign these publications at the threshold to the PR classes.

<div align="center">**PLEASE INSERT TABLE 1 HERE**</div>

*Counting items with lower citation rates*

In agreement with the proposal of Leydesdorff et al. (2011), Leydesdorff and Bornmann (2011) have applied the "counting rule that the number of items with lower citation rates than the item under study determines the percentile" (p.2137). This means that all the tied papers at a border are included in the lower

PR class.[1] Accordingly the factual threshold is always above the theoretical value as denoted in Table 1. In the present example this leads to factual interval boundaries between the PR classes at 50.36%, 76.44%, 90.05%, 95.15%, and 99.03%, see Table 1. Although these values are not too different from the theoretical thresholds, their distance varies enough to yield strong fluctuations in the percentages of publications which fall into the thus determined PR classes. In the present case notably the third PR class is somewhat underoccupied with 13.61% instead of 15% of the publications. Weighting the percentage of publications with the interval number $k$ yields the corresponding contribution to $R(6)$ as given in Table 1 so that the total score of all publications is 188.97% = 1.8897. This is somewhat below the theoretical expectation value of

$$R(6) = 50\% * 1 + 25\% * 2 + 15\% * 3 + 5\% * 4 + 4\% * 5 + 1\% * 6 = 191.00\ \% = 1.9100. \quad (1)$$

*Counting items with lower or equal citation rates*

Rousseau (2012) has suggested to include the item under study into the number of items to compare with. This effectively means that in the counting rule not only items with lower citation rates but items with lower or equal citations rates are taken into account. As a consequence, all the tied publications at the threshold are now always included into the higher PR class so that the factual threshold is always below the theoretical value, as denoted in Table 1. In the present example this leads to a strong deviation in particular for the first PR class which now comprises only 45.05% instead of 50% of all publications. This is counterbalanced by a high occupation of 29.29% instead of 25% for the second PR class. In summary, the contributions to $R(6)$ yield a total value of 1.9671 much higher than the theoretical expectation value (1).

*Middle of the uncertainty interval or average percentiles*

The difference between the factual thresholds in the two discussed approaches is the uncertainty interval defined by Leydesdorff (2012). Following Leydesdorff (2012) one could utilize the middle of that interval to categorize the publications at the threshold[2]: if the middle is below the threshold, then all tied papers are attributed to the lower PR class, if it is above the threshold, all papers would fall into the higher PR class. In principle there could be an ambiguity, because the middle of the interval might be exactly equal to the border, but this case is extremely unlikely, because it would not only require that the theoretical boundary corresponds to an integer value of publications, but also that the tied publications at the threshold are symmetrically distributed on both sides of the border.

Effectively, using the middle of the uncertainty interval is very similar to the approach of Pudovkin and Garfield (2009) who average the percentiles of the tied publications at the threshold. In order to avoid a distracting discussion about different ways of rounding percentages to integer percentile values, I have used the rational numbers $n/N$ for the $n$-th paper to determine the average quantile of the publications at the threshold as given in Table 1. The deviation from the middle of the uncertainty interval always amounts to exactly $1/2N$ and does not have an influence on the subsequent evaluation in the present example. The averaged quantile is now utilized according to Pudovkin et al. (2009) in the same way as Leydesdorff (2012) used the middle of the uncertainty interval described in the previous paragraph: if the average quantile is below/above the threshold, then all tied papers are attributed to the lower/upper PR class. The resulting factual threshold can therefore now lie below or above the PR class boundary, see Table 1. The deviations for the resulting percentages of publications in the various PR classes are somewhat smaller than in the previously discussed approaches and accordingly, after weighting the percentages with the interval number $k$, i.e., with the PR class number, the total score of $R(6) = 1.9128$ is rather close to the theoretical expectation value (1).

---

[1] The same method is utilized in a recent study comparing different universities (Bornmann, 2012a).

[2] For a single manuscript this middle of the uncertainty interval corresponds to the rule of Hazen (1914). It can be interpreted as the simplest linear interpolation between the boundaries of the uncertainty interval. Other proposals for the determination of percentiles often use slightly modified linear interpolation schemes (Gringorten, 1963; Cunnane, 1978; Harter, 1984). The quantile difference between these schemes for a single publication is smaller than $1/N$, but the main question of the present investigation remains open, namely how to treat tied papers.

*Average weights*

I have previously proposed (Schreiber, 2012b) to average not the percentiles of the tied papers, but rather to determine the individual weights and average these weights. The average (non-integer) weight should then be given to each of the tied papers. Effectively this means that the tied papers are shared by the two PR classes according to the number of tied papers below and above the border. Therefore the factual thresholds in this scheme are very close to the theoretical values, see Table 1. These factual thresholds can be obtained by rounding $p_k N$ to the next higher integer number, i.e., by the ceiling function, and then dividing by $N$. It is only due to the discretization that these values still deviate from the theoretical boundaries, and the deviation must always be smaller than $1/N$. Consequently the percentage of publications in each of the 6 PR classes is close to the theoretical distribution and the total score of $R(6) = 1.9090$ is also very close to 1.91.

I note that for this evaluation I have sorted the papers by number of citations and determined the quantile of each publication by the number of papers with a lower rank. Thus the factual threshold will always be (slightly) above the theoretical boundary value. If I would include the ranked paper in the paper count, then the factual threshold would be given by rounding $p_k N$ to the nearest lower integer number, i.e., by the floor function, divided by $N$. Therefore it would always be (slightly) smaller than the theoretical boundary value, namely exactly $1/N$ below the values given in Table 1. In this case one obtains a total score of 1.9111.

*Fractional scoring*

Although these deviations are agreeably small, I find them still irritating and thus the approach not really satisfactory. As mentioned in the introduction, the final solution is given by the fractional scoring scheme. In that method each publication is attributed a percentage interval, which for an individual publication corresponds to the uncertainty interval mentioned above. It is determined by ranking the publications according to their citations without regarding tied publications, i.e., giving the tied publications a random order. The $i$-th paper covers the interval from $(i-1)/N$ to $i/N$. Each publication will thus be attributed an interval of length $1/N$. The interval of a paper exactly at the threshold, in the present case that would be for example the 1187[th] paper at the 50% boundary, is fractionalized into the part (here one half of the 1187[th] paper) below and the part (its other half) above the boundary. These fractions need not be equal and in general at the other boundaries they will not be equal. They are then utilized to determine the average weight for this paper, and this is then used in the average over the weights of all the tied papers.

Equivalently, one can also first aggregate the intervals of all tied papers and thus start with one interval for all the tied papers at a boundary. The thus accumulated interval agrees with the uncertainty interval for tied papers discussed above. Now one has to fractionalize this interval into a part below and a part above the boundary (Waltman et al., 2013). The score is then determined by the weighted average of these two fractions. Because the summations involved may be exchanged, this weighted average is exactly equal to the average weight of the tied papers determined above where only the paper at the border was fractionalized.

Conceptually the latter approach of considering one comprehensive interval for all tied papers is more attractive, because it allows us to separate the attribution of the papers to different PR classes from the scoring. Now all tied papers are treated in the same manner, namely they are all counted fractionally in both PR classes below and above the border. One can visualize this fractionalization by considering the overlap between the uncertainty intervals for the publications and the percentile intervals for the PR classes (Waltman et al., 2013; Schreiber, 2012c). This fractional counting means that the shares of publications below and above the boundary correspond exactly to the theoretical values so that the perfect result of $R(6)$ = 1.9100 given in (1) is achieved for the total score as denoted in Table 1.

## Another example: Highly cited researchers

As a second example I use the citation data which I have harvested in July 2007 from the Web of Science for 8 highly cited physicists and which I have investigated in a different context previously (Schreiber, 2008b, 2010b). As above the search comprised only articles in the field of physics. Again the complete dataset is considered as the reference set and it is beyond the purpose of the present analysis to evaluate the individual scientists with respect to this reference set. The full dataset comprises 3354 publications with a total number of 279027 citations. Thus with 83.2 citations per publication the citation count is much higher than in my first example. Likewise the citation thresholds for the different PR classes are much higher, for example 22 citations are necessary to reach the 50% boundary, see Table 2. Nevertheless, there are 39 papers tied at this threshold, which means 1.16%. The factual thresholds for the above discussed 4 different counting rules are presented in Table 2 and in this case the percentages of publications in the 6 PR classes deviate not so strongly from the ideal distribution. Nevertheless, deviations of about 1% do occur and the total score ranges from $R(6)$ = 1.8962 to 1.9129. These deviations are much smaller than in the previous example, but again I find them irritating. In any case, they are unnecessary, because the fractional scoring scheme again avoids any deviation as shown in the last line of Table 2.

## Two further examples: Publications in a journal

Journal sets are recommended as reference sets (Bornmann, 2012). For this reason I present in Table 3 my evaluation of all articles that have appeared in the physics journal EPL in the years 2007 - 2010. I have determined the total number of 20997 citations for these 3203 papers from the Web of Science in January 2012 and utilized the data in a different context (Schreiber, 2012a). On average these papers have acquired 6.6 citations each and it is therefore not surprising that the 50% boundary is reached with 3 citations already, 7 citations are sufficient for the 75% boundary, and 90% of the papers have no more than 14 citations. Consequently, 10.02%, 3.47%, and 1.40% of all the publications can be found at these thresholds, many more than in the previous examples. Thus the factual thresholds deviate much more strongly from the boundary values than in the previous examples and therefore also the numbers of publications in the different PR classes are often far from the theoretical distribution. As a result the total score ranges from $R(6)$ = 1.8517 to $R(6)$ = 2.0066.

In this case the total score for the average-percentile approach is below the theoretical value (1) in contrast to the first two cases where it was above 1.9100. This shows, that the average-percentiles method of Pudovkin et al. (2009) as well as the middle-of-the-uncertainty interval procedure of Leydesdorff (2012) can deviate in either direction from the ideal value, while the other approaches are always leading to results either below (Leydesdorff and Bornmann, 2011) or above (Rousseau, 2012) $R(6)$ = 1.9100 for the total score (1). Fractionalizing the paper counts of the tied papers (Schreiber, 2012b) again yields a value very close to the ideal total score (1.9095, see Table 3, or 1.9110 if the ranked paper is included in the paper count), but only the fractional scoring scheme reproduces this theoretical expectation value (1) exactly again.

For the mentioned study (Schreiber, 2012a) I had also investigated all articles published in Europhysics Letters from 1999 until 2006. Europhysics Letters was rebranded EPL in 2007, so this is the same journal as discussed in the previous paragraph. Including the above analyzed 4-year period of EPL, altogether within these 12 years the journal has published 7553 papers which have been cited 87418 times as determined from the Web of Science in January 2012. Of course the older publications had much more time to be cited and therefore acquired more citations than those papers from the last 4 years evaluated in Table 3. On average now there are 11.6 citations per publication and this is also reflected in Table 4 where the numbers of citations at the PR class borders are about twice as large as in Table 3. Nevertheless the numbers of publications exactly at all these thresholds are somewhat larger. Due to the increased total number of papers, however, the percentages of publications at the thresholds are only about half as large as in Table 3. But the respective values are comparable with the percentages given in Table 1 for the first example, so are the deviations of the factual thresholds from the ideal boundaries and the deviations of the percentages of publications in the various PR classes from the theoretical distribution.

In this case the average quantiles of the publications at all thresholds are always above the boundary so that the tied papers are always attributed to the higher PR class. Thus for this dataset the average-quantile approach yields exactly the same results as the method in which the investigated items are included in the paper count for the determination of the percentile. Due to the larger number of publications the deviation of the averaged-weights approach is smaller than in the previous examples namely $R(6) = 1.9098$ (or 1.9192 if the ranked paper is included in the citation count). For completeness the last line in Table 4 again shows that fractional scoring leads to the perfect outcome (1).

## Concluding remarks

With 4 examples empirical evidence has been given that the determination of PR scores can be problematic not only for small publication sets. Due to the large numbers of tied publications which usually occur at the boundaries between the PR classes the differences between various methods for attributing the tied papers to different PR classes and thus for the calculation of the PR scores are indeed of more than "little or no practical consequence" in contrast to the claim made by Bornmann (2012). I found as many as 10% of the total number of publications at the 50% threshold in Table 3. Even at the 90% boundary there were as many as 1.40% of all publications. A similar observation was made by Waltman et al. (2013) where between 0.4% and 3.6% of all publications were found at the 90% boundary in 7 even larger datasets with up to 42749 publications. Thus the here presented values can be expected to be representative. Using even larger reference sets only means that more papers are tied at the thresholds. There is no reason to believe that the share of tied papers decreases. Thus the here discussed problems occur even in very large reference sets. Likewise, on average the individual publication sets, which are to be evaluated in comparison with the reference set, can also be expected to have a similar share of tied papers with the same citation numbers at the threshold unless these publication sets are very small. Thus is follows from the present investigation that the results of an evaluation will certainly be different if the various scoring schemes are applied. However, it remains an open question, whether this leads to significant changes in the ranking as long as the same reference set is used. Differences can be expected when different reference sets have to be utilized, e.g. in the comparison of publication sets for different fields (Waltman et al., 2013).

In conclusion, the differences between the various scoring methods are indeed relevant and do have practical consequences. Therefore fractional scoring is strongly recommended.

Table 1. Evaluation of the citation records of 26 researchers from the Institute of Physics at Chemnitz University of Technology, determining the contributions to the 6 PR classes in the $R(6)$ indicator.

| Percentile interval $k$ | 0 | 1 | 2 | 3 | 4 | 5 | 6 | total |
|---|---|---|---|---|---|---|---|---|
| Threshold $p_k$ | 0% | 50% | 75% | 90% | 95% | 99% | 100% | |
| No. citations at threshold | 0 | 4 | 12 | 25 | 43 | 104 | 457 | |
| No. pubs. below threshold | 0 | 1069 | 1764 | 2128 | 2254 | 2349 | 2372 | |
| No. pubs. at threshold | 477 | 126 | 50 | 9 | 4 | 1 | 1 | |
| No. pubs. above threshold | 1896 | 1178 | 559 | 236 | 115 | 23 | 0 | |
| % pubs. below threshold | 0.00 | 45.05 | 74.34 | 89.68 | 94.99 | 98.99 | 99.96 | |
| % pubs. at threshold | 21.10 | 5.31 | 2.10 | 0.37 | 0.16 | 0.04 | 0.04 | |
| % pubs. above threshold | 79.90 | 49.64 | 23.56 | 9.95 | 4.85 | 0.97 | 0.00 | |
| Factual threshold (LB) | 0.00 | 50.36 | 76.44 | 90.05 | 95.15 | 99.03 | 100.00 | |
| Factual threshold (R) | 0.00 | 45.05 | 74.34 | 89.68 | 94.99 | 98.99 | 100.00 | |
| Av. quantile of pubs. at threshold | | 47.72 | 75.41 | 89.89 | 95.09 | 99.03 | 100.00 | |
| Factual threshold (PG) | 0.00 | 50.36 | 74.34 | 90.05 | 94.99 | 98.99 | 100.00 | |
| Factual threshold (S) | 0.00 | 50.02 | 75.01 | 90.01 | 95.03 | 99.03 | 100.00 | |
| % pubs. in $k$-th PR class (LB) | | 50.36 | 26.08 | 13.61 | 5.10 | 3.88 | 0.97 | 100.00 |
| % pubs. in $k$-th PR class (R) | | 45.05 | 29.29 | 15.34 | 5.31 | 4.00 | 1.01 | 100.00 |
| % pubs. in $k$-th PR class (PG) | | 50.36 | 23.98 | 15.72 | 4.93 | 4.00 | 1.01 | 100.00 |
| % pubs. in $k$-th PR class (S) | | 50.02 | 24.99 | 15.00 | 5.01 | 4.00 | 0.97 | 100.00 |
| % pubs. in $k$-th PR class (WS) | | 50.00 | 25.00 | 15.00 | 5.00 | 4.00 | 1.00 | 100.00 |
| Contribution to $R(6)$ (LB) | | 50.36 | 52.16 | 40.83 | 20.40 | 19.40 | 5.82 | 188.97 |
| Contribution to $R(6)$ (R) | | 45.05 | 58.58 | 46.02 | 21.24 | 20.00 | 5.82 | 196.71 |
| Contribution to $R(6)$ (PG) | | 50.36 | 47.96 | 47.16 | 19.72 | 20.02 | 6.07 | 191.28 |
| Contribution to $R(6)$ (S) | | 50.02 | 49.98 | 45.01 | 20.06 | 20.02 | 5.82 | 190.90 |
| Contribution to $R(6)$ (WS) | | 50.00 | 50.00 | 45.00 | 20.00 | 20.00 | 6.00 | 191.00 |

Note. The abbreviations LB, R, PG, S, WS refer to the different scoring schemes by Leydesdorff and Bornmann (2011), Rousseau (2012), Pudovkin and Garfield (2009), Schreiber (2012b), Waltman and Schreiber (2012), respectively.

Table 2. Same as Table 1, but for 8 highly cited physicists; note that some lines from Table 1 are left out, because they are less important for the discussion.

| Percentile interval $k$ | 0 | 1 | 2 | 3 | 4 | 5 | 6 | total |
|---|---|---|---|---|---|---|---|---|
| Threshold $p_k$ | 0% | 50% | 75% | 90% | 95% | 99% | 100% | |
| No. citations at threshold | 0 | 22 | 63 | 171 | 354 | 1102 | 4192 | |
| No. pubs. below threshold | 0 | 1674 | 2512 | 3016 | 3186 | 3320 | 3353 | |
| No. pubs. at threshold | 384 | 39 | 12 | 3 | 1 | 1 | 1 | |
| No. pubs. above threshold | 2970 | 1641 | 830 | 335 | 167 | 33 | 0 | |
| % pubs. at threshold | 11.45 | 1.16 | 0.36 | 0.09 | 0.03 | 0.03 | 0.03 | |
| Factual threshold (LB) | 0.00 | 51.07 | 75.25 | 90.01 | 95.02 | 99.02 | 100.00 | |
| Factual threshold (R) | 0.00 | 49.91 | 74.90 | 89.92 | 94.99 | 98.99 | 100.00 | |
| Factual threshold (PG) | 0.00 | 49.91 | 74.90 | 90.01 | 94.99 | 98.99 | 100.00 | |
| Factual threshold (S) | 0.00 | 50.00 | 75.01 | 90.01 | 95.02 | 99.02 | 100.00 | |
| % pubs. in $k$-th PR class (LB) | | 51.07 | 24.18 | 14.76 | 5.01 | 4.00 | 0.98 | 100.00 |
| % pubs. in $k$-th PR class (R) | | 49.91 | 24.99 | 15.03 | 5.07 | 4.00 | 1.01 | 100.00 |
| % pubs. in $k$-th PR class (PG) | | 49.91 | 24.99 | 15.12 | 4.98 | 4.00 | 1.01 | 100.00 |
| % pubs. in $k$-th PR class (S) | | 50.00 | 25.01 | 15.00 | 5.01 | 4.00 | 0.98 | 100.00 |
| Contribution to $R(6)$ (LB) | | 51.07 | 48.36 | 44.28 | 20.04 | 19.98 | 5.90 | 189.62 |
| Contribution to $R(6)$ (R) | | 49.91 | 49.97 | 45.08 | 20.27 | 19.98 | 6.08 | 191.29 |
| Contribution to $R(6)$ (PG) | | 49.91 | 49.97 | 45.35 | 19.29 | 19.98 | 6.08 | 191.20 |
| Contribution to $R(6)$ (S) | | 50.00 | 50.03 | 44.99 | 20.04 | 19.98 | 5.90 | 190.94 |
| Contribution to $R(6)$ (WS) | | 50.00 | 50.00 | 45.00 | 20.00 | 20.00 | 6.00 | 191.00 |

Table 3. Same as Table 2, but for all publications in the physics journal EPL from 2007 until 2010.

| Percentile interval $k$ | 0 | 1 | 2 | 3 | 4 | 5 | 6 | total |
|---|---|---|---|---|---|---|---|---|
| Threshold $p_k$ | 0% | 50% | 75% | 90% | 95% | 99% | 100% | |
| No. citations at threshold | 0 | 3 | 7 | 14 | 20 | 45 | 444 | |
| No. pubs. below threshold | 0 | 1395 | 2320 | 2876 | 3029 | 3171 | 3202 | |
| No. pubs. at threshold | 508 | 321 | 111 | 45 | 17 | 2 | 1 | |
| No. pubs. above threshold | 2695 | 1487 | 772 | 282 | 157 | 30 | 0 | |
| % pubs. at threshold | 15.86 | 10.02 | 3.47 | 1.40 | 0.53 | 0.06 | 0.03 | |
| Factual threshold (LB) | 0.00 | 53.57 | 75.90 | 91.20 | 95.10 | 99.06 | 100.00 | |
| Factual threshold (R) | 0.00 | 43.55 | 72.43 | 89.79 | 94.57 | 99.00 | 100.00 | |
| Factual threshold (PG) | 0.00 | 53.57 | 75.90 | 89.79 | 95.10 | 99.00 | 100.00 | |
| Factual threshold (S) | 0.00 | 50.02 | 75.02 | 90.01 | 95.00 | 99.00 | 100.00 | |
| % pubs. in $k$-th PR class (LB) | | 53.57 | 22.32 | 15.30 | 3.90 | 3.97 | 0.94 | 100.00 |
| % pubs. in $k$-th PR class (R) | | 43.55 | 28.88 | 17.36 | 4.78 | 4.43 | 1.00 | 100.00 |
| % pubs. in $k$-th PR class (PG) | | 53.57 | 22.32 | 13.98 | 5.31 | 3.90 | 1.00 | 100.00 |
| % pubs. in $k$-th PR class (S) | | 50.02 | 25.01 | 14.99 | 5.00 | 4.00 | 1.00 | 100.00 |
| Contribution to $R(6)$ (LB) | | 53.57 | 44.65 | 45.89 | 15.61 | 19.83 | 5.62 | 185.17 |
| Contribution to $R(6)$ (R) | | 43.55 | 57.76 | 52.08 | 19.11 | 22.17 | 5.99 | 200.66 |
| Contribution to $R(6)$ (PG) | | 53.57 | 44.65 | 41.68 | 21.23 | 19.51 | 5.99 | 186.64 |
| Contribution to $R(6)$ (S) | | 50.02 | 50.02 | 44.96 | 19.98 | 19.98 | 5.99 | 190.95 |
| Contribution to $R(6)$ (WS) | | 50.00 | 50.00 | 45.00 | 20.00 | 20.00 | 6.00 | 191.00 |

Table 4. Same as Table 2, but for all publications in the physics journal Europhysics Letters / EPL from 1999 until 2010.

| Percentile interval $k$ | 0 | 1 | 2 | 3 | 4 | 5 | 6 | total |
|---|---|---|---|---|---|---|---|---|
| Threshold $p_k$ | 0% | 50% | 75% | 90% | 95% | 99% | 100% | |
| No. citations at threshold | 0 | 6 | 14 | 27 | 40 | 87 | 536 | |
| No. pubs. below threshold | 0 | 3673 | 5654 | 6795 | 7173 | 7473 | 7551 | |
| No. pubs. at threshold | 732 | 380 | 169 | 48 | 18 | 6 | 1 | |
| No. pubs. above threshold | 6820 | 3499 | 1729 | 709 | 361 | 73 | 0 | |
| % pubs. at threshold | 9.69 | 5.03 | 2.24 | 0.64 | 0.24 | 0.08 | 0.01 | |
| Factual threshold (LB) | 0.00 | 53.67 | 77.11 | 90.61 | 95.22 | 99.03 | 100.00 | |
| Factual threshold (R) | 0.00 | 48.64 | 74.87 | 89.98 | 94.98 | 98.95 | 100.00 | |
| Factual threshold (PG) | 0.00 | 48.64 | 74.87 | 89.98 | 94.98 | 98.95 | 100.00 | |
| Factual threshold (S) | 0.00 | 50.00 | 75.00 | 90.00 | 95.01 | 99.01 | 100.00 | |
| % pubs. in $k$-th PR class (LB) | | 53.67 | 23.44 | 13.51 | 4.61 | 3.81 | 0.97 | 100.00 |
| % pubs. in $k$-th PR class (R) | | 48.64 | 26.23 | 15.11 | 5.01 | 3.97 | 1.05 | 100.00 |
| % pubs. in $k$-th PR class (PG) | | 48.64 | 26.23 | 15.11 | 5.01 | 3.97 | 1.05 | 100.00 |
| % pubs. in $k$-th PR class (S) | | 50.00 | 25.00 | 15.00 | 5.01 | 4.00 | 0.99 | 100.00 |
| Contribution to $R(6)$ (LB) | | 53.67 | 46.88 | 40.52 | 18.43 | 19.07 | 5.80 | 184.36 |
| Contribution to $R(6)$ (R) | | 48.64 | 52.46 | 45.33 | 20.02 | 19.86 | 6.28 | 192.58 |
| Contribution to $R(6)$ (PG) | | 48.64 | 52.46 | 45.33 | 20.02 | 19.86 | 6.28 | 192.58 |
| Contribution to $R(6)$ (S) | | 50.00 | 50.00 | 45.01 | 20.02 | 19.99 | 5.96 | 190.98 |
| Contribution to $R(6)$ (WS) | | 50.00 | 50.00 | 45.00 | 20.00 | 20.00 | 6.00 | 191.00 |